\documentclass[useAMS,usenatbib]{mn2e}

\def\msun{\hbox{M$_\odot$}}

\def\t4{\hbox{t$_{\rm 4}$}}

\def\vesc{\hbox{V$_{\rm esc}$}}

\usepackage{graphicx}
\usepackage{natbib}
\usepackage{amssymb}
\usepackage{lscape}
\title[A lack of age spreads in young LMC clusters]{Constraints on possible age spreads within young massive clusters in the Large Magellanic Cloud}
\author[N. Bastian \& E. Silva-Villa]{ N. Bastian$^{1}$\thanks{NB: N.J.Bastian@ljmu.ac.uk} \& E. Silva-Villa$^{2}$\\
$^{1}$ Astrophysics Research Institute, Liverpool John Moores University, Egerton Wharf, Birkenhead, CH41 1LD, UK\\
$^{2}$ D\'epartement de Physique, de G\'enie Physique et d'Optique, and Centre de Recherche en Astrophysique du Qu\'ebec \\ (CRAQ), Universit\'e Laval, Qu\'ebec, QC G1V 0A6, Canada}
\begin{document}

\date{Accepted XXX. Received XXX; in original form XXX}

\pagerange{\pageref{firstpage}--\pageref{lastpage}} \pubyear{2013}

\maketitle

\label{firstpage}

\begin{abstract}
Recent studies have shown that the observed main-sequence turn-off (MSTO) in colour-magnitude diagrams of intermediate age (1-2~Gyr) clusters in the LMC are broader than would be nominally expected for a simple stellar population.  This has led to the suggestion that such clusters may host multiple stellar populations, with age spreads of 100-500~Myr.  However, at intermediate ages, spreads of this magnitude are difficult to discern and alternative explanations have been put forward (e.g., stellar rotation, interacting binaries).  A prediction of the age-spread scenario is that younger clusters in the LMC, with similar masses and radii, should also show significant age spreads.  In younger clusters (i.e., 40-300~Myr) such large age spreads should be readily apparent.  We present an analysis of the colour-magnitude diagrams of two massive young clusters in the LMC (NGC~1856 and NGC~1866) and show that neither have such large age spreads, in fact, both are consistent with a single burst of star-formation ($\sigma{\rm(age)} < 35$~Myr).  This leads us to conclude that either the intermediate age clusters in the LMC are somehow special or that the broadened MSTOs are not due to an age spread within the clusters.
\end{abstract}


\section{Introduction}

The traditional view of stellar clusters as simple stellar populations (SSP - i.e. all stars have the same age and metallicity within some small tolerance) has recently been called into question due to the discovery of multiple main sequences, horizontal branches, turn-offs, and sub-giant branches in the old globular ($>10$~Gyr) clusters in the Galaxy (e.g., Bedin et al.~2004).  However, it is currently unclear if this is a general property of massive stellar clusters, or if this feature is unique to the globular clusters.  High precision photometry of intermediate age ($1-2$~Gyr) clusters in the LMC has shown that while the main-sequence and giant branches are well described by an SSP model, the main sequence turn-off is broader than would be expected, even when photometric errors are taken into account (e.g., Mackey \& Broby Nielsen~2007, Milone et al.~2009, Goudfrooij et al.~2011a,b).  This broadening has been taken as evidence for a significant age spread within the clusters  ($>100-500$~Myr), which if true, could provide vital clues into how the multiple populations in globular clusters formed (e.g. Conroy \& Spergel~2011, Keller, Mackey, \& Da Costa~2011).

Alternatives explanations to age spreads have been put forward to explain the extended main sequence turnoff (eMSTO) seen in LMC clusters.  Yang et al.~(2011) have suggested that interacting binaries can create an eMSTO as well as a dual red clump, observed in some intermediate age clusters (Girardi et al.~2009).  Additionally, Bastian \& de Mink~(2009; see also Yang, Bi, \& Meng~2012) suggested that stellar rotation may be able to produce an eMSTO due to the change in luminosity and colour in rapidly rotating stars although this scenario has been called into question (Girardi et al.~2011).

If massive intermediate age clusters do host significant age spreads or multiple populations, this could provide a promising avenue to understand the multiple populations observed in globular clusters.  Indeed, there have been many recent works that have attempted to link the suggested age spread in the intermediate age LMC clusters with globular clusters (e.g., Conroy \& Spergel~2011, Keller et al.~2011).  These works have suggested that massive clusters ($\gtrsim10^4$\msun) clusters have deep enough potentials to retain material lost due to stellar evolution of the first generation of stars and potentially accrete new material from their surroundings.  If such material can be retained, a second (or multiple) populations of stars can be formed within the existing cluster.  This is similar to models put forward to explain the multiple populations within globular clusters (e.g., de Mink et al~2009, and references therein), although such models may require a larger population of first generation stars than can be accommodated by observations of the number of field stars within the host galaxy (Larsen, Strader, \& Brodie~2012).

Other studies have suggested that age spreads may be present in young ($<100$~Myr), massive ($>10^5$\msun) extragalactic clusters (e.g., Larsen et al.~2011).  These potential age spreads are significantly less in duration, being tens of Myr, instead of hundreds of Myr in the intermediate age LMC clusters.  However, the authors conclude that interacting binaries or other effects may explain the results without resorting to age spreads, especially as the estimated spread in ages appears to be correlated to the age of the cluster (Larsen et al.~2011).

\begin{figure}
\includegraphics[width=7.5cm]{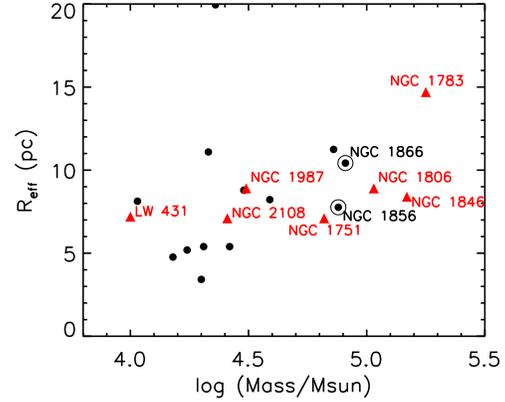}
\caption{The mass vs. effective radius for a sample of relatively young ($<500$~Myr) massive ($>10^4$\msun) clusters in the LMC (black dots), taken from the catalogue of McLaughlin \& van der Marel~(2005).  The two clusters discussed in the present paper are marked with large open circles.  Intermediate age clusters that show the eMSTO feature, from the sample of Goudfrooij et al.~(2011) are shown as filled (red) triangles.  Note that the clusters presented in the current work are more massive than many of the clusters showing the eMSTO feature and have similar effective radii.}
\label{fig:mass_reff}
\end{figure}

In the present work we look for significant age spreads in two massive clusters in the LMC, NGC~1856 and NGC~1866.  Both clusters are massive ($\sim10^5$\msun, see Table~\ref{tab:objects} for the basic cluster properties) and have properties similar to clusters showing the eMSTO feature (see Fig.~\ref{fig:mass_reff}).  If the age spreads in the intermediate age LMC clusters are real, we would expect to find similar age spreads in these two clusters.  Both clusters have been studied previously, in particular NGC~1866 has an extensive history of being used for calibration of stellar evolutionary models (see Rosenfield et al.~in prep, for an extensive review). Brocato et al.~(2003) derived an age of $\sim160-250$~Myr for NGC~1866 based on an analysis of the colour-magnitude diagram (CMD).  NGC~1856 has been most recently studied by Kerber et al. (2007) who derived an age of $\sim280$~Myr.  We will use the photometry presented in each of the above works to search for possible age spreads within these clusters.

Throughout this work we adopt the Padova isochrones (Bressan et al.~2012) at a metallicity of z=0.008, a distance modulus to the LMC of 18.50 (e.g., Marconi \& Clementini~2005), and a Salpeter~(1955) stellar initial mass function.


\begin{table*} 

  \begin{tabular}
    {lcccccc} ID& log (Age/yr) & log (Mass/\msun)&R$_{\rm eff}$ (pc)& R$_{\rm core}$ (pc) & V$_{\rm esc}$ (km/s)\\
    \hline 
NGC~1856 & 8.12 (8.45) & 4.88 & 7.8 & 1.7& 10.5\\
NGC~1866 & 8.12 (8.25) & 4.91 & 10.4 & 2.8 & 9.2\\
    \hline 
  \end{tabular}
\caption{Properties of the two LMC YMCs discussed in the present work.  All values are taken from McLaughlin \& van der Marel~(2005), adopting King~(1966) profile fits.  The number in parenthesis is the best fit age from the star formation history analysis presented in this work.  Note that V$_{\rm esc}$ would be a factor $\sim1.2-1.5$ higher if we applied the corrections of Goudfrooij et al.~(2011b) to estimate V$_{\rm esc}$ at an age of 10~Myr.}
\label{tab:objects}
\end{table*}

\section{NGC 1866: Hess diagrams and the star formation history}
\label{sec:1866}

NGC~1866 is a relatively young (100-200~Myr), massive cluster ($\sim10^5$\msun) in the LMC (e.g.~McLaughlin \& van der Marel~2005).  In order to constrain a possible age spread within this cluster we use {\em Hubble Space Telescope} - {\em Wide Field Planetary Camera 2} (HST-WFPC2) images taken in the {\it F555W} and {\it F814W} bands.  The data were presented in detail in Brocato et al.~(2003), who analysed the images to study the stellar luminosity function within the cluster in order to constrain the stellar mass function and the efficiency of core overshooting in stellar models.  In the present work we use their stellar photometric catalogues that have been transformed to the standard {\em VI} bands.

Here, we focus on the inner $4.8$~pc of the cluster, in order to avoid background contamination as much as possible.  No background subtraction was performed, as background stars will only increase the appearance of any potential age spreads that are present.

\begin{figure*}
\includegraphics[width=14cm]{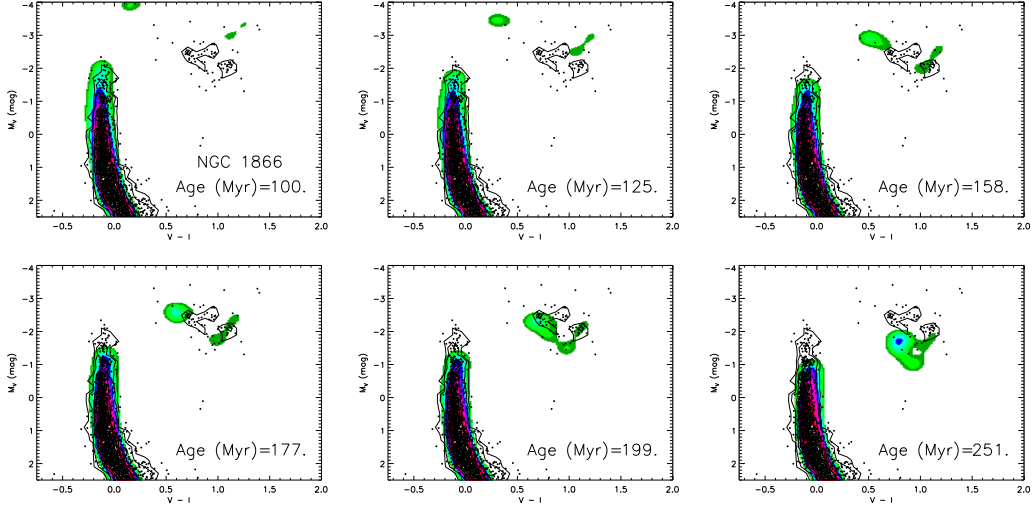}
\caption{Hess diagrams for NGC 1866.  The dots represent individual stars, the black contours show the binned data (i.e. the number of stars in a given colour/magnitude bin) and the filled (colour) contours represent the models at the age given in each panel.  Note the strong constraints given by the morphology/position of the helium burning stars.  The best fitting age is $\sim180$~Myr, with an acceptable range of $\sim30$~Myr on either side, based on the helium burning stars.  Note the handful of stars above the nominal main sequence for the best fit age.}
\label{fig:ngc1866_hess}
\end{figure*}


In Figure~\ref{fig:ngc1866_hess} we show the CMD of the observed stars (back points) along with contours (solid dark lines) representing the number density of stars at a given colour/magnitude.  The main sequence is clearly evident and the helium burning stars have two over-densities.  Additionally, in the six panels we show Hess diagrams adopting the Padova isochrones and our photometric errors for six different ages, ranging from 100 to 250~Myr.  We find $A_{\rm V}=0.15$, consistent with previous studies (e.g., Brocato et al.~2003).  Focussing on the position and morphology of the helium burning stars, we see that the best fit is obtained at $\sim180$~Myr.  Significant numbers of stars from younger ages can be excluded as they would produce helium burning star features that do not correspond to the observations.  Older populations can also be excluded due the lack of stars with similar colours, but fainter than, the observed helium burning stars.

One feature that should be noted is that the observed main sequence extends to brighter magnitudes than the predictions of the Hess diagram for the best fit age, by $\sim0.5$~mag.  These stars could be due to a younger generation, although evidence against this is seen in the lack of corresponding post-main sequence (i.e., helium burning) stars.  Alternatively, these stars could be due to binaries in the cluster and/or crowding in the images since the observations are focussed on the relatively dense core of the cluster.

\begin{figure}
\includegraphics[width=8cm]{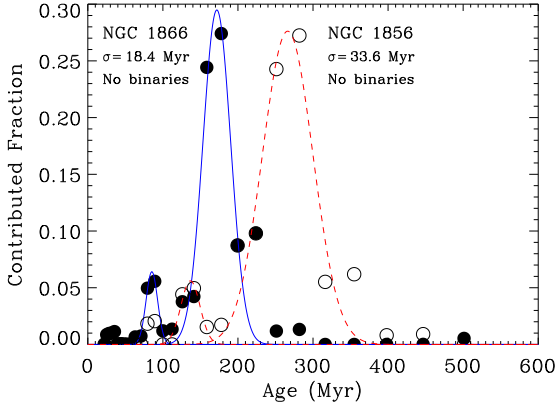}
\caption{The derived SFH of NGC 1866 (solid cirlces) and NGC~1856 (open circles), assuming no binary stars.  The points represent the estimations, while the solid (blue) and dashed (red) lines show Gaussian fits to the points for NGC~1866 and NGC~1856, respectively.  For NGC~1866 the main peak has a dispersion of $\sigma=18.4$~Myr and a peak at $180$~Myr, while for NGC~1856 we find a dispersion of 33.6~Myr and a peak at 280~Myr.  The dispersions are upper limits to the true age distribution.  The secondary peaks are driven by the handful of stars above the main sequence turn-off (see Figs.~\ref{fig:ngc1866_hess} \& \ref{fig:ngc1856_hess}), which are likely due to binaries and crowding effects in the central regions of the clusters, since no corresponding post-main sequence features are seen.}
\label{fig:ngc1866_sfh}
\end{figure}

In order to provide a more quantitative measure of possible age spreads within NGC~1866, we also fit the star-formation history (SFH) of the cluster using the observed CMD, Padova isochrones and the {\em FITSFH} code presented in Silva-Villa 
\& Larsen~(2010) and Larsen et al.~(2011). We restricted the fits to brighter than $M_{\rm V} = 1$ in order to avoid incompleteness issues.   We refer to the reader to the above works for details on the fitting procedure.  Photometric errors were included in the fits, and we chose boxes in the CMDs (regions within the CMDs to compare the observations with the isochrones) with the following limits: for NGC~1866 $-0.3 \le (V-I) \le 0.3$ and $-4.0 \le M_{V} \le 1.5$ for the main sequence and  $0.3 \le (V-I) \le 1.5$ and $-4.0 \le M_{V} \le 0.05$ for the post-main sequence.  We experimented with different limits and found that the results were not sensitive to the exact choice of colours/magnitudes over which to fit.  The resulting SFH is shown in Fig.~\ref{fig:ngc1866_sfh}, under the assumption that no binary stars are present, as solid circles.  The solid (blue) lines show two Gaussians fit to the data.  The main peak has a dispersion of $18.4$~Myr.  The secondary peak is due to the handful of stars above the nominal main sequence turnoff (for the best fitting age).  This peak can be reduced or removed by including the effects of binaries in the fitting procedure, although crowding effects are likely to be contributing as well.  However, the inclusion of binaries increases the dispersion of the main peak from 18~Myr to 30~Myr.  

From this we can see that any possible age spread is limited to 30~Myr of the peak age.  This is significantly less than the 100-500~Myr expected from observations of the intermediate age clusters in the LMC.

\section{NGC 1856: Hess diagrams and the star formation history}
\label{sec:1856}

Another of the most massive young ($<1$~Gyr) clusters in the LMC is NGC~1856.  Using the mass and size estimated in the survey of Macky \& Gilmore~(2003), Goudfrooij et al.~(2011b) suggest that this is the only cluster with an escape velocity, $\vesc$, larger than the empirically derived ``critical" escape velocity of $10$~km/s, hence that it should host multiple generations of stars within it.

\begin{figure*}
\includegraphics[width=18cm]{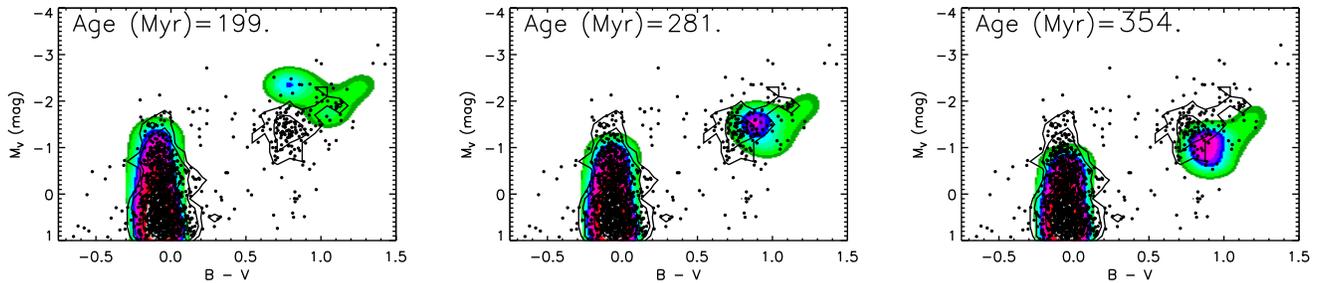}
\caption{Similar to Fig.~\ref{fig:ngc1866_hess} but now for NGC~1856.  Using the helium burning stars position and morphology, the best fit age is $\sim280$~Myr.  Due to the noisier data, the age spread constraint is less than in NGC~1866, with significant populations of stars with ages $\pm50$~Myr not consistent with the observations.  As in NGC~1866, there are a handful of stars brighter than the nominal main sequence turnoff, for the best fit age, presumably due to crowding and binaries, as no counterpart post-main sequence population is seen.}
\label{fig:ngc1856_hess}
\end{figure*}

We can test this suggestion by studying the CMD of NGC~1856 presented in Brocato et al.~(2001).  These authors present HST-WFPC2 F450W and F555W observations of the cluster, with the PC chip positioned near the cluster core.  We refer the interested reader to that paper for the observational details.  The CMD, uncleaned for field and foreground star contamination, is shown in Fig.~\ref{fig:ngc1856_hess}, with points representing individual stars and the solid (black) line contours show the number density of points at that colour/magnitude.  As in \S~\ref{sec:1866} we restrict our analysis to the central 4.8~pc to minimise contamination of the sample. We also show theoretical Hess diagrams for three different ages in the panels.

Using the position and morphology of the helium burning stars we can constrain the age and any age spread present within the cluster.  From Fig.~\ref{fig:ngc1856_hess} we see that the helium burning stars are best reproduced by the $280$~Myr isochrone.  However, due to the lower signal-to-noise of the data, the constraint on the possible age spread within the cluster (based on helium burning star morphology) is weaker than in the case of NGC~1866.  We can rule out significant populations of stars with ages less/more than $50$~Myr than the best fit isochrone.

As in NGC~1866, there are a handful of stars above the nominal main sequence turnoff.  These are presumably binary stars and/or stars affected by crowding due to the high surface density of stars in the central parts of the cluster.  The possibility of these stars representing a younger population is low, due to the lack of corresponding post-main sequence stars.

We have also estimated the SFH of NGC~1856 using the same method as in \S~\ref{sec:1866}.   For this cluster we adopted fitting boxes with limits of $-0.3 \le (B-V) \le 0.3$ and $-4.0 \le M_{V} \le 0.0$ for the main sequence and $0.3 \le (V-I) \le 1.5$ and $-4.0 \le M_{V} \le 0.0$ for the post-main sequence.  We find $A_{\rm V} = 0.8$, consistent with previous studies (e.g., Kerber et al.~2007), and the estimated SFH is shown in Fig.~\ref{fig:ngc1866_sfh} as open circles.  The dashed lines show two Gaussian fits to the estimated distribution.  We find a best fit age of 280~Myr and a dispersion of $\sim35$~Myr.  As in the case of NGC~1866, the secondary peak at younger ages is due to the handful of stars above the nominal main sequence of the best fit age, presumably due to binaries and crowding effects.

\section{Discussion and Conclusions}

We have tested the assertion that large age spreads ($>100$~Myr) are a common feature of massive clusters.  These suggestions were based on previous studies of massive ($>10^4\msun$), intermediate age (1-3~Gyr) clusters that showed that the observed main sequence turnoff (MSTO) was broader than predicted by simple stellar population, which could be explained by an extended star-formation history.  Alternatively, it has been suggested that extended MSTOs (eMSTO) could be due to stellar rotation or interacting binaries.  If the eMSTOs are due to an age spread within the cluster, then it would be expected that younger clusters (ages $<500$~Myr) with similar properties (mass and radius) should show a clear age spread ($100-500$~Myr) within them.  Such an age spread would be easily detectable through the analysis of their CMDs.  We note that the eMSTO clusters are generally better described by a continuous distribution of ages, rather than a bimodal distribution (e.g., Goudfrooij et al. 2011b).

\begin{figure}
\includegraphics[width=8cm]{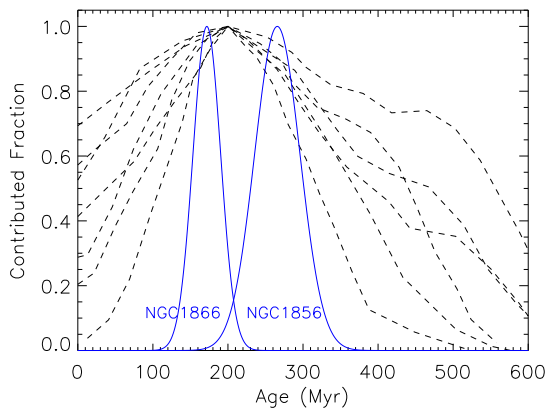}
\caption{A comparison between the estimated SFH of NGC~1856 and NGC~1866 in the current work (solid blue lines) and the expectations from seven intermediate age clusters in Goudfrooij et al.~(2011b) shifted to a reference age of 200~Myr.  All clusters have been normalised so that the peak contribution is unity.  The distributions of the two clusters presented in the current work are significantly narrower, hence do not conform to the expectation if age spreads were a common feature in massive clusters.}
\label{fig:comparison}
\end{figure}

Using CMDs based on HST imaging of two massive young clusters in the LMC (NGC~1856 and NGC~1866; see Table~\ref{tab:objects} for their properties) we find that both are consistent with a single burst of star formation, and place upper limits to possible age spreads within them of $\sigma=18$~Myr and $\sigma=35$~Myr for NGC~1866 and NGC~1856, respectively (fitting the derived SFH with a Gaussian).  These are upper limits to the true dispersion, as crowding, contamination,  and differential extinction has not been taken into account (additionally the errors in the observations introduce some minimum width to the best fit SFH).  The lack of an age spread in each of these clusters is consistent with that found for massive ($10^4 - 10^5$\msun) young ($<10$~Myr) clusters in the Galaxy, such as NGC~3603 and Westerlund 1 (e.g.~Kudryavtseva et al.~2012).  In Fig.~\ref{fig:comparison} we show Gaussian fits to the SFH of both clusters, as well as the expected distribution based on the inferred SFH of seven intermediate age clusters presented in Goudfrooij et al.~(2011b), which have been shifted to an age of 200~Myr for comparison.  The two clusters presented in the current work clearly do not match the inferred SFH of the intermediate age clusters.

The lack of age spreads within the two young clusters in the current work can be interpreted in two ways.  Either the intermediate age clusters with eMSTOs are fundamentally different than the clusters presented here in some intrinsic property (although their masses and radii are comparable) or the eMSTO feature observed in these clusters is not due to an age spread.  Alternative explanations for the eMSTOs have been put forward in the literature, namely interacting binary stars (Yang et al.~2011) and stellar rotation (Bastian \& de Mink~2009).  Girardi et al.~(2011) have called the rotation scenario into question, by correctly pointing out that the models presented by Bastian \& de Mink were not isochrones, as the influence of rotation on stellar lifetimes was not taken into account.  When Girardi et al. did correct for the changing lifetimes, the spread in the modelled MSTO decreased.  However, we note that the rotation rate used by Girardi et al. was rather low (using the mean of the Royer et al.~(2007) rotational distribution and not taking into account the significant numbers of more rapidly rotating stars) and that their models did not show any difference between the rotating and non-rotating stars on the main sequence, although rotation should affect stellar structure and be observable on the main sequence (e.g., Brott et al.~2011; Ekstr{\"o}m et al.~2012).


Both of the clusters discussed here are well above the $10^4$\msun\ limit suggested by Conroy and Spergel~(2011) in order for clusters to host multiple populations.  Both clusters have an escape velocity of $>10$~km/s (after corrections for mass loss and change of radius, using the method discussed in Goudfrooij et al.~2011b) which is larger than many of the clusters that display the eMSTO feature, and the clusters discussed here have similar properties as the eMSTO clusters.  

The observations presented here, of a lack of an age spread within two young massive clusters, suggests that the eMSTO feature in intermediate age clusters is not due to age spreads.  This is also consistent with the lack of abundance spreads within intermediate age clusters (e.g., Mucciarelli et al. 2007), and suggests that the eMSTO feature is not related to the multiple stellar populations seen in globular clusters.  Additionally, the observed colour spread in the eMSTO feature is similar in all intermediate age clusters observed to date, resulting in an age spread that is proportional to the age of the cluster, suggesting that other effects may be causing the eMSTO feature.


In a future work, we will present an analysis of the observed CMDs of another $\sim15$ young LMC/SMC clusters.  Additionally, we will present modelling of each of the clusters, including the two clusters presented here, that involves a more realistic treatment of the errors as well as the effects of crowding and binarity.  Through this method we will be able to place the tightest possible constraints on possible age spreads within these clusters.  


\section*{Acknowledgments}
We thank Enzo Brocato for providing the NGC~1866 data in electronic format and  are grateful to Ben Davies, Selma de Mink, Dan Weisz, Paul Gooudfrooij and Phil Rosenfield for insightful discussions.  NB  is funded by a University Research Fellowship from the Royal Society.  ES-V is a postdoctoral fellow supported by the Centre de Recherche en Astrophysique du Qu\'ebec (CRAQ).



\begin{thebibliography}{99}

\bibitem[Bastian 
\& de Mink(2009)]{2009MNRAS.398L..11B} Bastian, N., \& de Mink, S.~E.\ 2009, MNRAS, 398, L11 

\bibitem[Bedin et al.(2004)]{2004ApJ...605L.125B} Bedin, L.~R., Piotto, G., 
Anderson, J., et al.\ 2004, ApJL, 605, L125 

\bibitem[Bressan et al.(2012)]{2012MNRAS.427..127B} Bressan, A., Marigo, 
P., Girardi, L., et al.\ 2012, MRNAS, 427, 127 

\bibitem[Brocato et 
al.(2001)]{2001A&A...374..523B} Brocato, E., Di Carlo, E., \& Menna, G.\ 2001, A\&A, 374, 523 

\bibitem[Brocato et al.(2003)]{2003AJ....125.3111B} Brocato, E., 
Castellani, V., Di Carlo, E., Raimondo, G., 
\& Walker, A.~R.\ 2003, AJ, 125, 3111 

\bibitem[Brott et 
al.(2011)]{2011A&A...530A.116B} Brott, I., Evans, C.~J., Hunter, I., et al.\ 2011, A\&A, 530, A116 



\bibitem[Conroy 
\& Spergel(2011)]{2011ApJ...726...36C} Conroy, C., \& Spergel, D.~N.\ 2011, ApJ, 726, 36 

\bibitem[de Mink et 
al.(2009)]{2009A&A...507L...1D} de Mink, S.~E., Pols, O.~R., Langer, N., \& Izzard, R.~G.\ 2009, A\&A, 507, L1 

\bibitem[Ekstr{\"o}m et 
al.(2012)]{2012A&A...537A.146E} Ekstr{\"o}m, S., Georgy, C., Eggenberger, P., et al.\ 2012, A\&A, 537, A146 


\bibitem[Girardi et al.(2009)]{2009MNRAS.394L..74G} Girardi, L., Rubele, 
S., \& Kerber, L.\ 2009, MNRAS, 394, L74 

\bibitem[Girardi et al.(2011)]{2011MNRAS.412L.103G} Girardi, L., 
Eggenberger, P., \& Miglio, A.\ 2011, MNRAS, 412, L103 

\bibitem[Goudfrooij et al.(2011)]{2011ApJ...737....3G} Goudfrooij, P., 
Puzia, T.~H., Kozhurina-Platais, V., \& Chandar, R.\ 2011a, ApJ, 737, 3 

\bibitem[Goudfrooij et al.(2011)]{2011ApJ...737....4G} Goudfrooij, P., 
Puzia, T.~H., Chandar, R., \& Kozhurina-Platais, V.\ 2011b, ApJ, 737, 4 

\bibitem[Keller et al.(2011)]{2011ApJ...731...22K} Keller, S.~C., Mackey, 
A.~D., \& Da Costa, G.~S.\ 2011, ApJ, 731, 22 

\bibitem[Kerber et 
al.(2007)]{2007A&A...462..139K} Kerber, L.~O., Santiago, B.~X., \& Brocato, E.\ 2007, A\&A, 462, 139 

\bibitem[King(1966)]{1966AJ.....71...64K} King, I.~R.\ 1966, AJ, 71, 64 

\bibitem[Kudryavtseva et al.(2012)]{2012ApJ...750L..44K} Kudryavtseva, N., 
Brandner, W., Gennaro, M., et al.\ 2012, ApJL, 750, L44 

\bibitem[Larsen et 
al.(2011)]{2011A&A...532A.147L} Larsen, S.~S., de Mink, S.~E., Eldridge, J.~J., et al.\ 2011, A\&A, 532, A147 

\bibitem[Larsen et 
al.(2012)]{2012A&A...544L..14L} Larsen, S.~S., Strader, J., \& Brodie, J.~P.\ 2012, A\&A, 544, L14 


\bibitem[Mackey 
\& Gilmore(2003)]{2003MNRAS.338...85M} Mackey, A.~D., \& Gilmore, G.~F.\ 2003, MNRAS, 338, 85 

\bibitem[Mackey 
\& Broby Nielsen(2007)]{2007MNRAS.379..151M} Mackey, A.~D., \& Broby Nielsen, P.\ 2007, MNRAS, 379, 151 

\bibitem[Marconi 
\& Clementini(2005)]{2005AJ....129.2257M} Marconi, M., \& Clementini, G.\ 2005, AJ, 129, 2257 


\bibitem[McLaughlin 
\& van der Marel(2005)]{2005ApJS..161..304M} McLaughlin, D.~E., \& van der Marel, R.~P.\ 2005, ApJS, 161, 304 

\bibitem[Milone et 
al.(2009)]{2009A&A...497..755M} Milone, A.~P., Bedin, L.~R., Piotto, G., \& Anderson, J.\ 2009, A\&A, 497, 755 

\bibitem[Mucciarelli et al.(2008)]{2008AJ....136..375M} Mucciarelli, A., 
Carretta, E., Origlia, L., \& Ferraro, F.~R.\ 2008, AJ, 136, 375 


\bibitem[Royer et 
al.(2007)]{2007A&A...463..671R} Royer, F., Zorec, J., \& G{\'o}mez, A.~E.\ 2007, A\&A, 463, 671 

\bibitem[Salpeter(1955)]{1955ApJ...121..161S} Salpeter, E.~E.\ 1955, ApJ, 
121, 161 


\bibitem[Silva-Villa 
\& Larsen(2010)]{2010A&A...516A..10S} Silva-Villa, E., \& Larsen, S.~S.\ 2010, A\&A, 516, A10 



\bibitem[Yang et al.(2011)]{2011ApJ...731L..37Y} Yang, W., Meng, X., Bi, 
S., et al.\ 2011, ApJL, 731, L37 

\bibitem[Yang et al.(2012)]{2012arXiv1212.1511Y} Yang, W., Bi, S., 
\& Meng, X.\ 2012, RAA, submitted (arXiv:1212.1511)


\end{thebibliography}
\end{document}